\documentclass[twocolumn,showpacs,preprintnumbers,pre]{revtex4}
\usepackage{amssymb}
\usepackage{amsmath}
\usepackage{graphicx}
\usepackage{dcolumn}
\usepackage{bm}

\begin{document}

\title{Revisiting the self-contained quantum refrigerator in the strong
coupling regime}
\email{quaninformation@sina.com; ycs@dlut.edu.cn}
\author{Chang-shui Yu}
\author{Qing-yao Zhu}
\affiliation{School of Physics and Optoelectronic Technology, Dalian University of
Technology, Dalian 116024, P. R. China}
\date{\today }

\begin{abstract}
We revisit the self-contained quantum refrigerator in the strong internal coupling regime by employing quantum optical master equation. It is shown that the strong internal coupling reduces the cooling ability of the refrigerator. In contrast to the weak coupling case,  the strong internal coupling could lead to quite different and even converse thermodynamic behaviors. \end{abstract}
\maketitle

\section{Introduction.}

Thermodynamics is one of the four pillars of theoretical physics and provides us with an essential way to study the thermodynamic process such as heat
engine which can be dated back to Carnot [1]. When we consider the physical nature down to the quantum level, quantum thermodynamics which is the
intersection of thermodynamics and quantum mechanics, provides a new approach
to investigate the microscopic physics. Quantum thermodynamics has
attracted more and more interests such as in Refs. [2-4] and the references therein. In particular,
quantum heat engine has been extensively studied  [5-12]. It was shown that
quantum heat engine has the remarkable similarity to the classical engines which obey
macroscopic dynamics and Carnot efficiency has been a well established limit for
some quantum heat engines [13-17]. A lot of works have been done
especially related to quantum analogues of Carnot engines [18-22], whilst some other
cycles such as Otto cycles [23-26] and Brownian motions [27] are also covered
with considerable progress. All above provide microscopic alternatives to test
the fundamental laws of thermodynamics and deepen our understanding of quantum
thermodynamics.

Recently, the concept of the self-contained quantum refrigerator has been
raised for the questions about the fundamental limitation on the size of
thermal machines and their relevant topics [28-32]. It is shown that the
`self-contained' means 1) all degrees of freedom of the refrigerator are taken
into account; 2) no external source of work is allowed; and 3) in particular, time-dependent Hamiltonians or prescribed unitary transformations are not
allowed. However, the key in their model is that they required the
interaction (the coupling) between their three qubits was weak enough, but
the coupling and the decay rate are on the same order. In other words, the
self-contained refrigerator works in the regime of weak internal coupling.
Since the three-qubit interaction is the vital driving mechanism for the
cooling, could a strong internal interaction (coupling) provide a more
effective power?

In this paper, we revisit the same model proposed in Ref. [28] in the strong internal coupling regime. We employ the quantum optical
master equation to study the steady-state heat currents and the cooling
efficiency. As the main result, we find that the strong internal coupling
plays a negative role in the cooling ability. The thermodynamic properties of
such a model could also be different from and even opposite to those in the weak internal
coupling regime. In addition, it is shown that
our results will be consistent with those in Ref. [28] (the weak internal
coupling) if we reduce the internal coupling strength, although our master
equation, in principle, is only suitable for the strong internal coupling. This implies that the validity of the application
of the quantum master equation deserves our further consideration. This paper is organized as follows. In Sec.
II, we briefly introduce the interacting mechanism of the refrigerator and derive the
master equation. In Sec. III, we present our main results and make some necessary analysis. The
conclusion is obtained finally.

\section{The model and the master equation}
The refrigerator we considered here is made up of three atoms denoted, respectively,
by $R$, $C$ and $H$. The free Hamiltonian of the three-atom system is
given by
\begin{equation}
H_{0}=H_{R}+H_{C}+H_{H},
\end{equation}%
where $H_{\mu}=\frac{\omega _{\mu}}{2}\sigma _{\mu }^{z}$, $\omega _{\mu}$, $\mu =R
$, $C$ and $H$, is the transition frequency of Atom $\mu $, and $\sigma
^{z}=\left\vert e\right\rangle \left\langle e\right\vert -\left\vert
g\right\rangle \left\langle g\right\vert $ with $\left\vert e\right\rangle$ and $\left\vert g\right\rangle$ denoting the excited state and the ground
state. In particular, in order to guarantee the
resonant interaction, it is required that $\omega _{R}=\omega _{H}+\omega
_{C}$. Suppose that the interaction of the three atoms is described by the Hamiltonian $H_I$:
\begin{equation}
H_{I}=g\left( \sigma _{H}^{+}\sigma _{R}^{-}\sigma _{C}^{+}+\sigma
_{H}^{-}\sigma _{R}^{+}\sigma _{C}^{-}\right)
\end{equation}%
with $g$ the coupling constant and $\sigma ^{+}=\left\vert e\right\rangle
\left\langle g\right\vert $ and $\sigma ^{-}=\left\vert g\right\rangle
\left\langle e\right\vert $, the Hamiltonian of the closed system
reads
\begin{equation}
H_{S}=H_{0}+H_{I}.
\end{equation}
Here we set the Planck constant and Boltzmann's constant to be unit, i.e., $%
\hbar =k_{B}=1$. In the framework of self-contained refrigerator
[28,30], all the atoms should interact with a reservoir respectively, instead of a real working source. So we let Atom $H$ be connected with a hot reservoir with the
temperature denoted by $T_{H}$, Atom $R$ be in contact with a "room" reservoir with
temperature $T_{R}$ and Atom $C$ interact with a cold reservoir with
temperature $T_{C}$. Thus It is naturally implied that $T_{H}>T_{R}>T_{C}$. Here we assume that all the reservoirs consist of infinite harmonic
oscillators with closely spaced frequencies $\nu _{\mu k}$ and annihilation operators  $b_{\mu k}$. Note that the subscript $\mu$ marks the atom which the corresponding reservoir interacts with. Thus one can write the total Hamiltonian of the open system as
\begin{equation}
H=H_{S}+\sum_{\mu }\left( H_{\mu 0}+H_{\mu }\right) ,
\end{equation}%
where $H_{\mu 0}=\sum\limits_{k}\nu _{\mu
k}b_{\mu k}^{\dagger }b_{\mu k}$ is the free Hamiltonian of the $%
\mu $th reservoir, and
\begin{equation}
H_{\mu }=\sum\limits_{k}f_{\mu k}(b_{\mu k}^{\dagger }\sigma _{\mu
}^{-}+b_{\mu k}\sigma _{\mu }^{+})
\end{equation}%
with $f_{\mu k}$ denoting the coupling constant,  describes the interaction between the $\mu$th atom and its thermal
reservoir. From Eq. (4), i.e., the total Hamiltonian, in principle,  one can obtain all the dynamics of the
refrigerator and the reservoirs. To do so, we have to derive a master equation that governs the evolution of the
system of interests. Next, we will follow the standard procedure [33,34] to find such a master equation.

Since the refrigerator (excluding the reservoirs) is a composite quantum system, the first step is to diagonalize the refrigerator
Hamiltonian $H_{S}$. It is shown that  the diagonalized $H_{S}$ can be written as $H_{S}=\sum \epsilon
_{i}\left\vert \lambda _{i}\right\rangle \left\langle \lambda
_{i}\right\vert $, where the eigenvalues are given by
\begin{equation}
\left[ \epsilon _{1},\epsilon _{2},\cdots ,\epsilon _{8}\right] =[\omega
_{R},\omega _{H},g,-\omega _{C},\omega _{C},-g,-\omega _{H},-\omega _{R}],
\end{equation}%
and $\left\vert \lambda _{i}\right\rangle $ denote the corresponding
eigenvectors with the concrete form omitted here. In $H_S$ representation, the
Hamiltonian  $H$ can be rewritten as%
\begin{equation}
H=\sum_{i=1}^{8}\epsilon _{i}\left\vert \lambda _{i}\right\rangle
\left\langle \lambda _{i}\right\vert +\sum_{\mu ,j}\left( H_{\mu 0}+H_{\mu
j}^{\prime }\right) ,
\end{equation}%
where
\begin{equation}
H_{\mu j}^{\prime }=\sum\limits_{k}f_{\mu k}(b_{\mu k}^{\dagger }V_{\mu
j}(w_{\mu j})+b_{\mu k}V_{\mu j}^{\dag }(w_{\mu j}))
\end{equation}%
with $V_{\mu j}(w_{\mu j}))$ denoting the eigenoperators of the refrigerator
Hamiltonian $H_{S}$ such that $\left[ H_{S},V_{\mu j}(w_{\mu j})\right]=-w_{\mu j}V_{\mu j}(w_{\mu j})$ and $w_{\mu j} $ standing for the eigenfrequency.
 In particular,  $V_{\mu j}(\nu _{j})$ can be explicitly
given as follows.
\begin{eqnarray}
V_{11} &=&\left\vert \lambda _{5}\right\rangle \left\langle \lambda
_{1}\right\vert +\left\vert \lambda _{8}\right\rangle \left\langle \lambda
_{4}\right\vert ,w_{11}=\omega _{H}, \\
V_{12} &=&\frac{1}{\sqrt{2}}\left( \left\vert \lambda _{3}\right\rangle
\left\langle \lambda _{2}\right\vert +\left\vert \lambda _{7}\right\rangle
\left\langle \lambda _{6}\right\vert \right) ,w_{12}=\omega _{H}-g, \\
V_{13} &=&\frac{1}{\sqrt{2}}\left( \left\vert \lambda _{7}\right\rangle
\left\langle \lambda _{3}\right\vert -\left\vert \lambda _{6}\right\rangle
\left\langle \lambda _{2}\right\vert \right) ,w_{13}=\omega _{H}+g, \\
V_{21} &=&\frac{1}{\sqrt{2}}\left( \left\vert \lambda _{3}\right\rangle
\left\langle \lambda _{1}\right\vert -\left\vert \lambda _{8}\right\rangle
\left\langle \lambda _{6}\right\vert \right) ,w_{21}=\omega _{R}-g, \\
V_{22} &=&\left\vert \lambda _{4}\right\rangle \left\langle \lambda
_{2}\right\vert +\left\vert \lambda _{7}\right\rangle \left\langle \lambda
_{5}\right\vert ,w_{22}=\omega _{R}, \\
V_{23} &=&\frac{1}{\sqrt{2}}\left( \left\vert \lambda _{8}\right\rangle
\left\langle \lambda _{3}\right\vert +\left\vert \lambda _{6}\right\rangle
\left\langle \lambda _{1}\right\vert \right) ,w_{23}=\omega _{R}+g, \\
V_{31} &=&\frac{1}{\sqrt{2}}\left( \left\vert \lambda _{3}\right\rangle
\left\langle \lambda _{5}\right\vert +\left\vert \lambda _{4}\right\rangle
\left\langle \lambda _{6}\right\vert \right) ,w_{31}=\omega _{C}-g, \\
V_{32} &=&\frac{1}{\sqrt{2}}\left( \left\vert \lambda _{4}\right\rangle
\left\langle \lambda _{3}\right\vert -\left\vert \lambda _{6}\right\rangle
\left\langle \lambda _{5}\right\vert \right) ,w_{32}=\omega _{C}+g, \\
V_{33} &=&\frac{1}{\sqrt{2}}\left( \left\vert \lambda _{2}\right\rangle
\left\langle \lambda _{1}\right\vert +\left\vert \lambda _{8}\right\rangle
\left\langle \lambda _{7}\right\vert \right) ,w_{33}=\omega _{C},
\end{eqnarray}%
where $w_{\mu j}>0$ is implied, otherwise, $V_{\mu j}=V_{\mu j}^{\dag }$.
Suppose that the system and their reservoirs are initially separable and the initial states of the reservoirs are the thermal equilibrium states. In particular, we assume that the coupling between the system and the reservoirs is weak enough. Based on the
Born-Markovian approximations, one can derive the master equation as
\begin{equation}
\dot{\rho}=\mathcal{L}_{C}[\rho ]+\mathcal{L}_{R}[\rho ]+\mathcal{L}%
_{H}[\rho ],
\end{equation}%
where the dissipators read
\begin{eqnarray}
&&\mathcal{L}_{\mu }[\rho ]=\sum_{j}J_{\mu }\left( -w_{\mu j}\right)
\left[ 2V_{\mu j}\left( w_{\mu
j}\right) \rho V_{\mu j}^{\dag }\left( w_{\mu j}\right) \right.\notag\\
&&-\left.V_{\mu j}^{\dag
}\left( w_{\mu j}\right) V_{\mu j}\rho \left( w_{\mu j}\right) -\rho V_{\mu
j}^{\dag }\left( w_{\mu j}\right) V_{\mu j}\left( w_{\mu j}\right) \right]
\notag \\
&&+J_{\mu }\left( w_{\mu j}\right)
\left[ 2V_{\mu j}^{\dag }\left( w_{\mu j}\right) \rho V_{\mu j}\left( w_{\mu
j}\right)\right.\notag\\
&&-\left.V_{\mu j}\left( w_{\mu j}\right) V_{\mu j}^{\dag }\left( w_{\mu
j}\right) \rho -\rho V_{\mu j}\left( w_{\mu j}\right) V_{\mu j}^{\dag
}\left( w_{\mu j}\right) \right] .\quad
\end{eqnarray}
 The spectral density in Eq. (19) is given by
\begin{eqnarray}
J_{\mu }\left( w_{\mu j}\right)  &=&\gamma _{\mu }\left( w_{\mu j}\right)
\bar{n}\left( w_{\mu j}\right) , \\
J_{\mu }\left( -w_{\mu j}\right)  &=&\gamma _{\mu }\left( w_{\mu j}\right)
\left[ \bar{n}\left( w_{\mu j}\right) +1\right] ,
\end{eqnarray}%
where $\bar{n}\left( w_{\mu j}\right)$ is  the average photon number which depends on the temperature of
the reservoir, i.e.,
\begin{equation}
\bar{n}\left( w_{\mu j}\right) =\frac{1}{e^{\frac{w_{\mu j}}{T_{\mu }}%
}-1}.
\end{equation}%
Here we suppose that $\gamma _{\mu }\left( w_{\mu j}\right) =\gamma _{\mu }$
is frequency-independent for simplicity. In addition, we employed the rotating wave approximation, which implies
$\gamma _{\mu }<<\left\vert\omega _{\mu}-\omega _{\nu}\pm 2g\right\vert, g $. This condition requires that the master equation is only suitable for the large $g$. However, so far there hasn't been an explicit constraint on to what degree
$g$ is larger than $\gamma_\mu$ [34,35].
\begin{figure}[tbp]
\includegraphics[width=0.8\columnwidth]{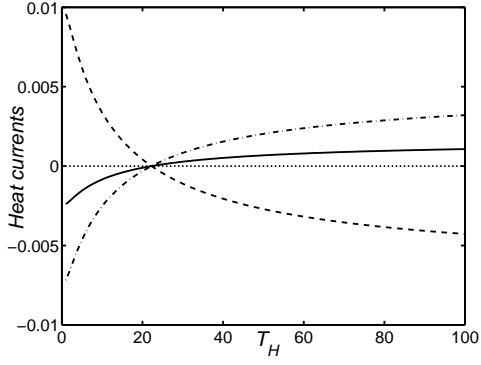}
\caption{The heat currents $\dot{Q}_\protect\mu[J/s]$ versus $T_H[K]$ in weak
coupling regime. The solid line, the dashed line and the dash-dotted line
correspond to $\dot{Q}_C$, $\dot{Q}_H$ and $\dot{Q}_R$, respectively. Here $%
g=0.001\protect\omega_H$. In particular, we set $\protect\gamma=0.001\protect%
\omega_H$ throughout the paper. }
\label{1}
\end{figure}

\section{Results and discussions}

In order to study the thermodynamical behavior of the stationary state, we will find the stationary-state solution $\rho ^{S}$ of the master equation given by Eq. (18).  To do so,  we let $\rho ^{S}$ have the vanishing derivative on $t$, i.e.,
\begin{equation}
\dot{\rho}^{S}=0.
\end{equation}%
Thus we will arrive at the following equations
\begin{eqnarray}
M\left\vert \rho \right\rangle &=&0, \\
\rho _{ij}^{S} &=&0,i\neq j,
\end{eqnarray}%
where $\left\vert \rho \right\rangle =[\rho _{11}^{S},\rho _{22}^{S},\rho
_{33}^{S},\rho _{44}^{S},\rho _{55}^{S},\rho _{66}^{S},\rho _{77}^{S},\rho
_{88}^{S}]^{T}$ is the vector made up of the diagonal entries of the
stationary density matrix $\rho ^{S}$, and
\begin{equation}
M=\sum\limits_{\mu =1}^{3}M_{\mu }.
\end{equation}%
In order to give the explicit expression for $M_\mu$, we first define some new quantities $m_{ij},i,j=1,2,3$ as
\begin{eqnarray}
m_{11} &=&2\mathbf{J}_{11}\otimes \left( \mathbf{1}_{+}\otimes \mathbf{1}_{+}+%
\mathbf{1}_{-}\otimes \mathbf{1}_{-}\right) , \\
m_{12} &=&\mathbf{1}\otimes C_{23}\left( \mathbf{J}_{12}\otimes \mathbf{1}%
_{-}\right) C_{23}^{\dag }, \\
m_{13} &=&\mathbf{J}_{13}\otimes \left( \mathbf{1}_{+}\otimes \mathbf{1}_{-}+%
\mathbf{1}_{-}\otimes \mathbf{1}_{+}\right) ,\\
m_{21} &=&\left( \mathbf{1}_{+}\otimes\mathbf{J}_{21}\otimes  \mathbf{1}_{+}+%
\mathbf{1}_{-}\otimes \mathbf{J}_{21}\otimes\mathbf{1}_{-}\right) , \\
m_{22} &=&2\left( \mathbf{1}_{+}\otimes\mathbf{J}_{22}\otimes  \mathbf{1}_{-}+%
\mathbf{1}_{-}\otimes \mathbf{J}_{22}\otimes\mathbf{1}_{+}\right), \\
m_{23} &=& C_{13}\left(\mathbf{J}_{23}\otimes  \mathbf{1}\otimes \mathbf{1}_{+}\right)C_{13}^\dagger ,\\
m_{31} &=&C_{21}\left(\mathbf{1}_{-}\otimes\mathbf{J}_{31}\otimes   \mathbf{1}\right)C_{21}^\dagger  , \\
m_{32} &=&\left( \mathbf{1}_{+}\otimes \mathbf{1}_{-}+%
\mathbf{1}_{-}\otimes \mathbf{1}_{+}\right)\otimes\mathbf{J}_{32} , \\
m_{33} &=&2\left( \mathbf{1}_{+}\otimes \mathbf{1}_{+}+%
\mathbf{1}_{-}\otimes \mathbf{1}_{-}\right)\otimes \mathbf{J}_{33},
\end{eqnarray}%
where
\begin{equation}
\mathbf{1=}\left(
\begin{array}{cc}
1 & 0 \\
0 & 1%
\end{array}%
\right) ,\mathbf{1}_{\pm }\mathbf{=}\frac{\mathbf{1}\pm \sigma _{z}}{2}
\end{equation}%
and $C_{jk}, j,k=1,2,3$ denotes the control-not gate with $j$ standing for the control qubit and $k$
representing the target qubit. For example,
\begin{equation}
C_{12}=\left(\mathbf{1}\oplus \sigma _{x}\right)\otimes \mathbf{1}.
\end{equation}%
In addition, $\mathbf{J}_{\mu j }$ in Eqs. (27-35) is a matrix with its entries corresponding to
the spectral density. It can be explicitly represented by
\begin{equation}
\mathbf{J}_{\mu j }=\left(
\begin{array}{cc}
-J_{\mu }\left( -w_{\mu j}\right) & J_{\mu }\left( w_{\mu j}\right) \\
J_{\mu }\left( -w_{\mu j}\right) & -J_{\mu }\left( w_{\mu j}\right)%
\end{array}%
\right) .
\end{equation}%
Based on Eqs. (27-35), $M_{\mu} $ can be explicitly written by
\begin{equation}
M_\mu=\sum_j m_{\mu j}.
\end{equation}
$M_\mu$ apparently includes three terms which are related to three atoms respectively.
Using the definition of the heat current [34,36],  we can find that the
heat current subject to $\mu $th reservoir reads
\begin{equation}
\dot{Q}_{\mu }=Tr\left\{ H_{S}\mathcal{L}_{\mu }\text{[}\rho ^{S}\text{]}%
\right\} =\left\langle \epsilon \right\vert M_{\mu }\left\vert \rho
\right\rangle .
\end{equation}%
It is obvious that $\dot{Q}_{\mu }$ corresponding to $M_\mu$ is uniquely determined by the steady state $\left\vert\rho\right\rangle$. It is fortunate that $\dot{Q}_{\mu }$ can be explicitly calculated, because Eq. (23) can be analytically solved. However, the concrete form
of $\left\vert \rho \right\rangle $ is so tedious that we cannot write it
here. Therefore,  in the following part, we will have to give a numerical analysis  based on
the analytical $\left\vert \rho \right\rangle$ (even though it is not given here).
\begin{figure}[tbp]
\includegraphics[width=0.8\columnwidth]{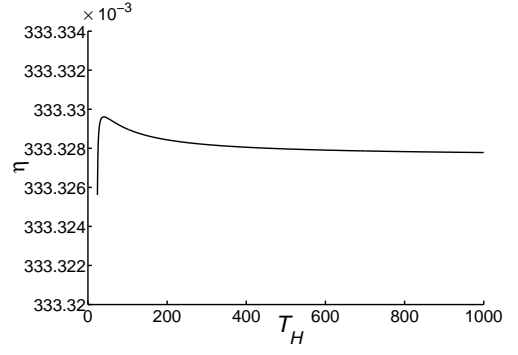}
\caption{ The efficiency $\eta$ of the refrigerator versus $T_H[K]$ in the weak
coupling regime. The efficiency changes slightly and it can be
considered to be almost invariant within a good approximation, which can also be
supported by Fig. 5.}
\label{1}
\end{figure}

At first, we would like to consider the weak coupling case, i.e., $g\sim\gamma_\mu$%
. Based on Eq. (40), we plot the heat currents in Fig. 1. Here we suppose $\omega
_{H}=3$ and $\omega _{C}=1$, so $\omega _{R}=4$. In addition, we let the room temperature be
$21K$ and the temperature of the cold reservoir be $18K$  (Of course, if the other parameters are chosen, one will
get the similar results). When the
temperature of the hot reservoir is low, the heat will flow into the cold
reservoir. So the cold atom $C$ is heated. However, with the temperature of
the hot reservoir increasing, one can find that all the heat currents will
become zero simultaneously when $T_{H}=T_{v}=\frac{\omega _{H}}{\frac{\omega _{R}%
}{T_{R}}-\frac{\omega _{C}}{T_{C}}}\simeq 22.24K$ (so long as the coupling $%
g $ and the decay rate $\gamma $ are small enough.). This virtual temperature $T_v$ is just
consistent with Ref. [32] which is closely related to Ref. [28]. In addition, one can
also see that the heat currents are increasing with the increase of $T_{H}$. That
is, the thermodynamic machine works as a refrigerator. An obvious feature is
that the heat currents subject to the hot reservoir and the cold reservoir
have the same direction (sign) and the sign is determined by the virtual
temperature $T_{v}$. However, if $\frac{\omega _{R}}{T_{R}}=\frac{\omega _{C}%
}{T_{C}}$, one will find that no matter how large $T_{H}$ is, $\dot{Q}_{C}$
is always less than zero.
In addition, we also consider the efficiency of the quantum refrigerator, which is illustrated in Fig. 2.
 Here the efficiency $\eta$ is defined by $\eta=\frac{\dot{Q}_{C}}{\dot{Q}_{H}}$ which was deeply studied in Ref. [30]. In a simple way, it can be understood as that, by extracting heat (current) $\dot{Q}_{H}$ from the hot reservoir, we are able to extract heat (current) $\dot{Q}_C$ from the cold reservoir whilst dumping heat (flow) $\dot{Q}_{R}$ into the reservoir $R$. It was also shown that $\eta$ for the self-contained refrigerator in Ref. [28] was given by $\frac{\omega _{C}}{\omega _{H}}$. Take the current parameters into account, it should be $\eta=\frac{1}{3}$. From our Fig. 2, at the first glance, the efficiency seems to have a peak
somewhere. But one can further find that to some acceptable
approximation, $\eta$ can be considered to be invariant on $T_{H}$ and just equal to $\frac{1}{3}$. The peak will be explained in the next part. All
these show that in the weak coupling regime, the treatment with respect to
the quantum optical master equation (QOME) has the well consistency with the previous
results given in Ref. [28]. \textit{This implies that the QOME could not be sensitive to the rotating wave approximation corresponding $g>>\gamma_\mu$ in this case, which could lead to that the QOME is valid here. }
\begin{figure}[tbp]
\includegraphics[width=0.8\columnwidth]{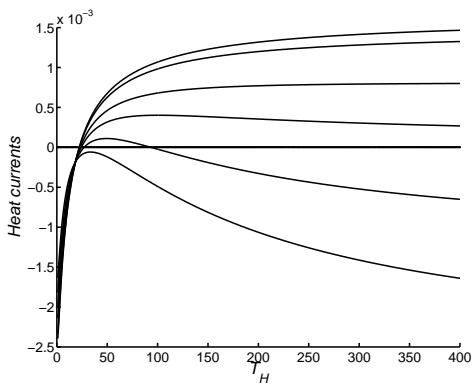}
\caption{The heat currents $\dot{Q}_C[J/s]$ versus $T_H[K]$ for different coupling
constants. From the top to the bottom, $g=0.001\protect\omega_H,0.1\protect%
\omega_H,0.2\protect\omega_H,0.25\protect\omega_H,0.3\protect\omega_H,0.35%
\protect\omega_H$. The straight line means zero heat current.}
\label{1}
\end{figure}\begin{figure}[tbp]
\includegraphics[width=0.8\columnwidth]{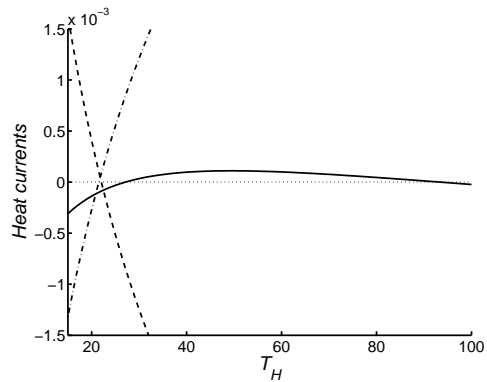}
\caption{ The heat currents versus $T_H[J/s]$ in the strong coupling regime. Here
$g=0.3\protect\omega_H$ and $\protect\gamma=0.001\protect\omega_H$. The
dotted line corresponds to the zero heat current and the other lines denote
the same heat currents as Fig. 1. $\dot{Q}_C$ is first pushed to the
positive direction and then is suppressed back to the negative direction. }
\label{1}
\end{figure}
\begin{figure}[tbp]
\includegraphics[width=0.8\columnwidth]{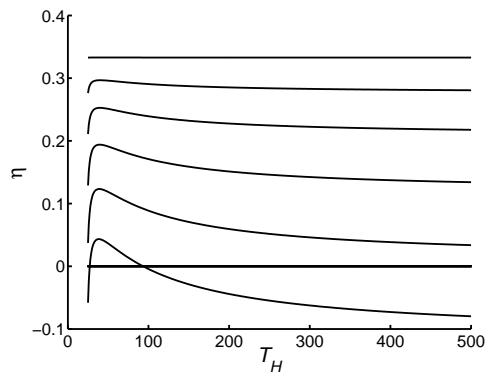}
\caption{The efficiency $\protect\eta$
versus $T_H[K]$ with different coupling constants. The lower straight line
corresponds to the zero efficiency. From the top to the bottom, the lines
correspond to $g=0.001\protect\omega_H,0.1\protect\omega_H,0.15\protect%
\omega_H,0.2\protect\omega_H,0.25\protect\omega_H,0.3\protect\omega_H$. In
particular, $\protect\eta= const.$ within acceptable approximations for $g=0.001\protect\omega_H$, which is
consistent to Fig. 2. }
\label{1}
\end{figure}
\begin{figure}[tbp]
\includegraphics[width=0.8\columnwidth]{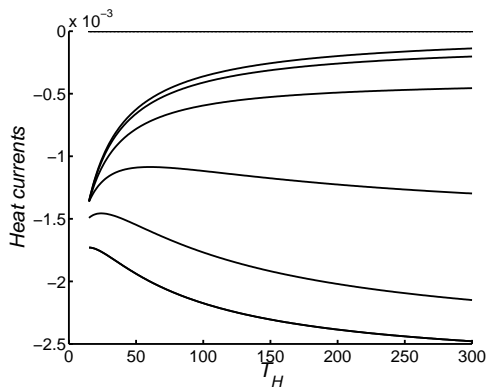}
\caption{The heat currents $\dot{Q}_C[J/s]$ versus $T_H[K]$ for different coupling
constants. Here we let $T_C=10K$ and $T_R=40K$ in order to satisfy $\frac{%
\protect\omega_C}{T_C}=\frac{\protect\omega_R}{T_R}=\frac{1}{10}$. The upper
straight line means zero heat current. From up to down, the lines correspond
to $g=0.001\protect\omega_H,0.1\protect\omega_H,0.2\protect\omega_H,0.3%
\protect\omega_H,0.4\protect\omega_H,0.5\protect\omega_H$, respectively. }
\label{1}
\end{figure}

Now let's turn to our main results i.e.,  $g>>\gamma_\mu$. To find the influence of
the coupling strength $g$, we keep $\omega_\mu$ and $%
\gamma_\mu$ invariant and plot the heat currents in Fig. 3 with different $g$.
One will immediately see that the large $g$ directly leads to the
suppression of the heat current $\dot{Q}_{C}$. Compared with the case of
weak coupling, the high temperature $T_{H}$ could have the negative role in
the cooling of the cold atom. It is obvious that the atom C cannot be cooled
if the coupling strength $g$ is too large, which is opposite to the case of weak internal coupling regime. In particular, given $T_{C}$, $T_{R}$
and all the frequencies, one will see that the cooling only happens within
some range of $T_{H}$, which has also been shown in Fig. 4. Thus the direct
conclusion is that the strong coupling is not beneficial to the cooling from
the point of refrigerator of view. In addition, one can also find that the
heat currents don't meet at a single point (temperature), which is quite
different from the weak coupling case. The heat currents don't change their
direction simultaneously. In particular, the heat current $\dot{Q}_{C}$
seems not to be directly relevant to the virtual temperature $T_{v}$. The
machine becomes a refrigerator only when $\dot{Q}_{C}>0$ where one will find
$T_{H}\simeq 27.25K\neq 22.24K=T_{v}$. As a refrigerator, the efficiency
depends on the coupling constant $g$. The numerical results are given in Fig.
5. It is shown that the efficiency will become larger if $g$ becomes less.
It will arrive at a constant efficiency ($\eta=\frac{\dot{Q}_{C}}{\dot{Q%
}_{H}}=\frac{\omega _{C}}{\omega _{H}}=\frac{1}{3}$) when it reaches the
weak coupling limit. However, the efficiency will change with $T_{H}$ if it
is still in the case of strong coupling. The peak of the efficiency mainly results from the suppression of cooling induced by the strong coupling. In particular, the suppression becomes strong for large $T_H$. When the reservoir $H$ is hot enough, the reservoir could be heated instead of cooled, which can be obviously found for $g=0.3\omega_H$. When the internal coupling become weak, the suppression will be weakened. If it is weak enough, the suppression won't be so apparent that the peak can
be neglected to some good approximation, which is just illustrated in Fig. 2. When $\frac{\omega _{R}}{T_{R}}=%
\frac{\omega _{C}}{T_{C}}$, one can also find that no matter what the
coupling constant is, it is impossible to make a refrigerator. However, from
a different angle,  we can find that in
the weak coupling limit, $%
\dot{Q}_{C}$ is reduced if we increase $T_{H}$. On the contrary, when the coupling is strong, $%
\dot{Q}_{C}$ become large with $T_{H}$ increasing. This is shown in Fig. 6.

\section{Conclusions}

In summary, we have revisited the self-contained refrigerator in the strong internal coupling regime by employing the quantum optical master equation. We find that the strong internal coupling reduces the cooling ability. In particular, in this regime, the considered machine demonstrates quite different (and even converse) thermodynamic behaviors compared with that in Ref. [28]. In addition, we find that the quantum optical master equation provides the consistent results with the Ref. [28] in the weak internal coupling regime, even though the rotating wave approximation, in principle, does not allow the weak internal coupling. This could shed new light on the validity of the master equation.

This work was supported by the National Natural Science Foundation of China,
under Grant No.11375036 and 11175033, and the Xinghai Scholar Cultivation Plan.

\end{document}